\documentstyle[11pt,newpasp,twoside,epsf]{article}
\def\edcomment#1{\iffalse\marginpar{\raggedright\sl#1\/}\else\relax\fi}
\reversemarginpar

\newcommand{\ergs}{\ifmmode {\rm ergs\,s}^{-1} \else ergs\,s$^{-1}$\fi}
\newcommand{\hb}{H\,$\beta$}

\newcommand{\civ}{C\,{\sc iv}}

\newcommand{\mbh}{$M_{\rm BH}$}
\newcommand{\lbol}{$L_{\rm bol}$}
\newcommand{\lol}{$L_{\rm bol}/L_{\rm Edd}$}
\newcommand{\Msol}{\mbox{$M_{\odot}$}}
\newcommand{\lsim}{\stackrel{\scriptscriptstyle <}{\scriptstyle {}_\sim}}
\newcommand{\gsim}{\stackrel{\scriptscriptstyle >}{\scriptstyle {}_\sim}}
\newcommand{\et}{\mbox{et~al.\ }}
\newcommand{\eg}{\mbox{e.g.,}\ }
\newcommand{\ie}{\mbox{i.e.,}\ }

\begin{document}
\title{Evidence for Early or Efficient Black-Hole Growth}
\author{M. Vestergaard}
\affil{Department of Astronomy, The Ohio State University, 140\,West\,18th Avenue, Columbus, OH\,43210}

\begin{abstract}
Mass estimates, based on scaling relationships, are presented of central 
black holes in luminous quasars at a range of redshifts ($z < 0.5$, 
$1.2 \lsim z \lsim 6.3$). The data show that very massive ($\gsim 10^9$\Msol) 
black holes appear already at $z \approx 6$, indicating that they form very 
early or very fast.
\end{abstract}

\section{Introduction}

Recent advances that allow virial black-hole masses in nearby quiescent and 
active galaxies to be measured have opened a new era: with mass measurements
we can address issues of the physics of black-hole evolution and its
connection with galaxy evolution.
Some obvious first questions to ask are what is the typical mass of black holes in 
AGNs at high redshift, and is this different from those of lower-$z$ AGNs? 
Also, does the typical mass of luminous quasars change beyond the redshift 
($z \approx 3$) at which the co-moving space density of quasars drops 
dramatically? A study addressing these issues (Vestergaard 2004, hereafter Paper~I)
is summarized below.

\section{Mass Estimates} 

As summarized by Vestergaard at this meeting, obtaining direct virial 
black-hole mass (\mbh) measurements of large samples of distant quasars is 
very time-consuming and is, hence, impractical. Instead, \mbh{} {\it estimates} 
can be more easily obtained with scaling relationships and measurements of 
broad-line width and continuum luminosity in a single-epoch UV or optical 
spectrum (\eg Wandel, Peterson, \& Malkan 1999; Vestergaard 2002; McLure 
\& Jarvis 2002).  The mass estimates presented here were obtained with such 
scaling relationships (Paper~I). 

But is it reasonable to assume that scaling relations, even those based on 
high-ionization lines, are valid for all AGNs and quasars? 
While the efficacy of the scaling relations is discussed in more detail in Paper~I,
a few key points are worth emphasizing here. 
From the demonstrated virial relationships (Peterson \& Wandel 1999, 2000;
Onken \& Peterson 2002) for the broad emission lines, including
the high-ionization lines, we know that for a given object the distance to the
line-emitting gas and its velocity dispersion each scale between the emission lines:
for example, $v^2_{\rm FWHM}$(\hb) $\cdot \tau$(\hb) 
$\propto v^2_{\rm FWHM}$(\civ) $\cdot \tau$(\civ), where $\tau$ is the light 
travel time delay between the variations in the ionizing continuum and responding
line emission. 
Therefore, 
any broad emission-line width (for which the virial relationship holds) can be used 
as a measure of the velocity dispersion, $v$(line), at the distance of this 
line-emitting gas, $R$(line) $ = c \tau$(line). In addition, since all four 
AGNs that can be 
tested exhibit this virial relationship, it is fair to assume it is universal
in spite of the small number of AGNs on which this conclusion is based.
Moreover, for the reasonable assumption that the broad-line region is 
photoionized one can show that
if the radius -- luminosity relationship did not apply to the luminous and
high-redshift quasars, then their
spectra would look very different than they actually do:
composite quasar spectra grouped by luminosity and redshift are surprisingly
similar at least up to redshifts of $\sim$4.5 (\eg Dietrich \et 2002). Also, the 
application of the radius -- luminosity relation to the most luminous (high-redshift)
quasars is only an extrapolation of at most 1.5 dex beyond the luminosity range 
of 4 dex over which the relationship is defined (Kaspi \et 2000). Thus, there 
are no immediate
concerns that the radius -- luminosity relationship \mbox{should not apply 
to typical luminous, distant quasars as well.}

\begin{figure}[t]
\epsfxsize=6cm
\plottwo{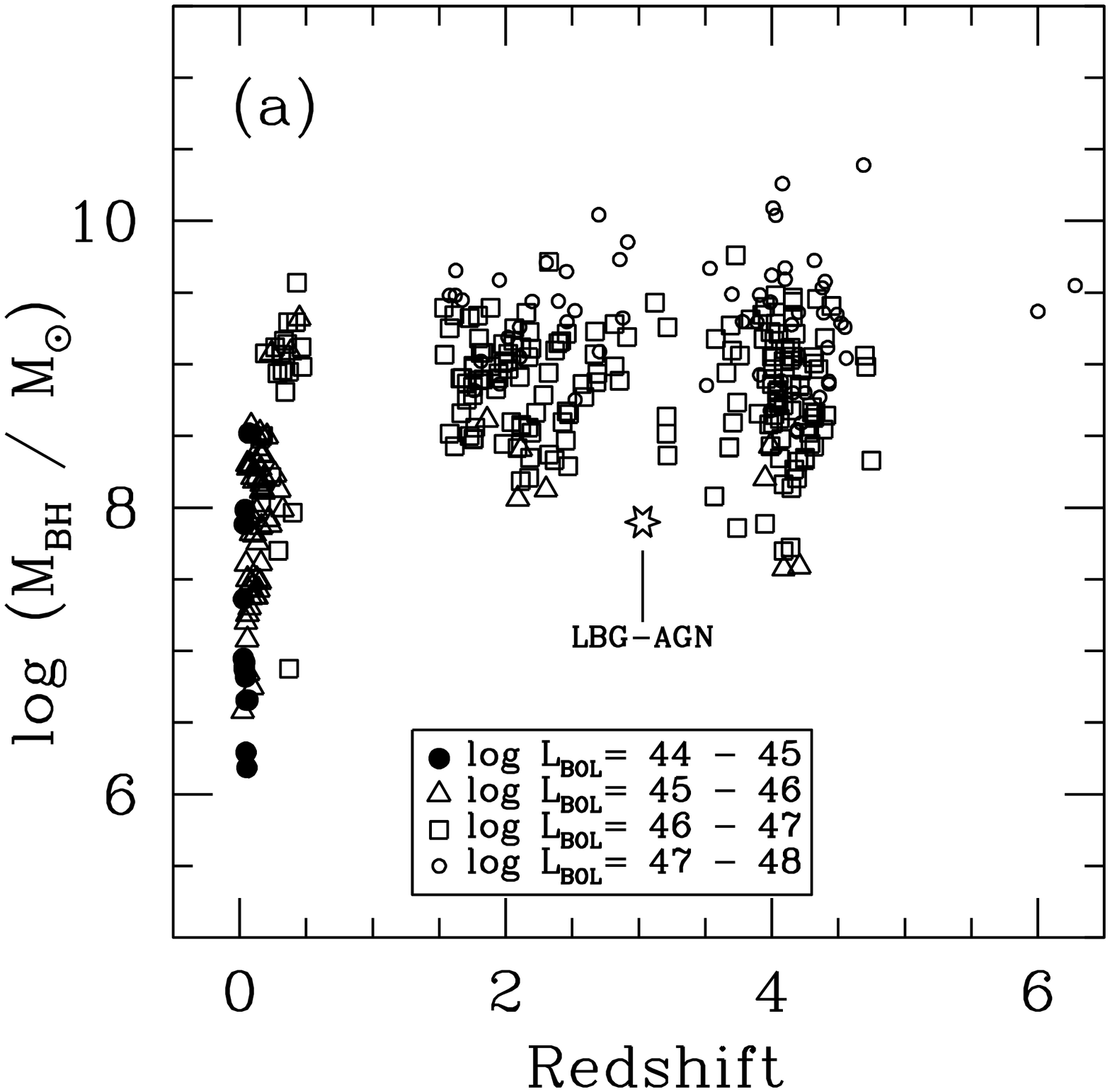}{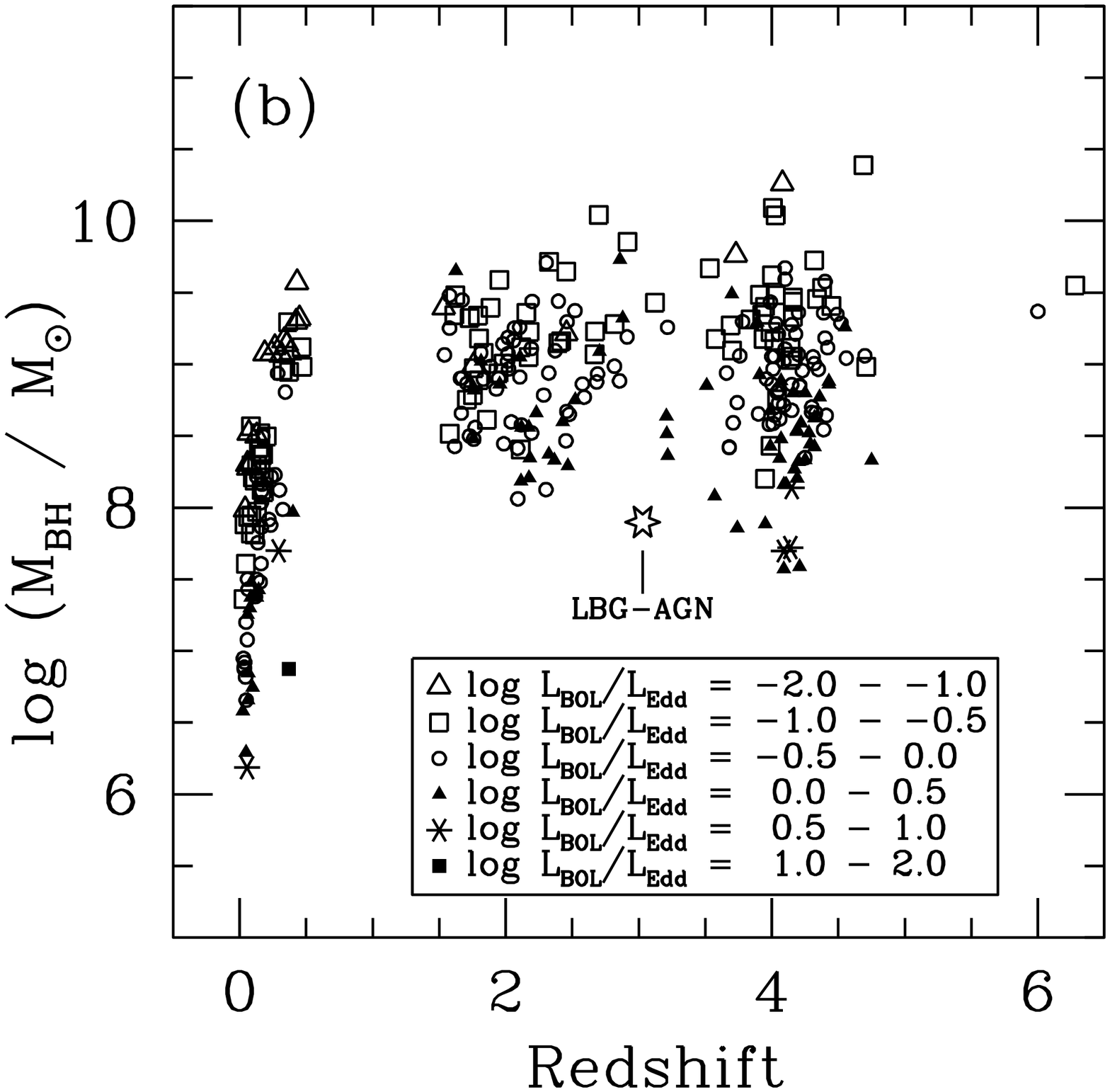}
\caption{Distribution of black-hole mass, \mbh, with redshift is shown binned
in ($a$) bolometric luminosity, \lbol{}, and in ($b$) the Eddington luminosity
ratio, \lol, of the samples described in \S~3. Errorbars are omitted for 
clarity. The dominant uncertainty is the statistical uncertainty in the scaling 
relations of a factor 2.5 to 3 relative to the reverberation mapping \mbox{to which 
the scaling relations are calibrated.} 
}
\end{figure}

\begin{figure}[t]
\plottwo{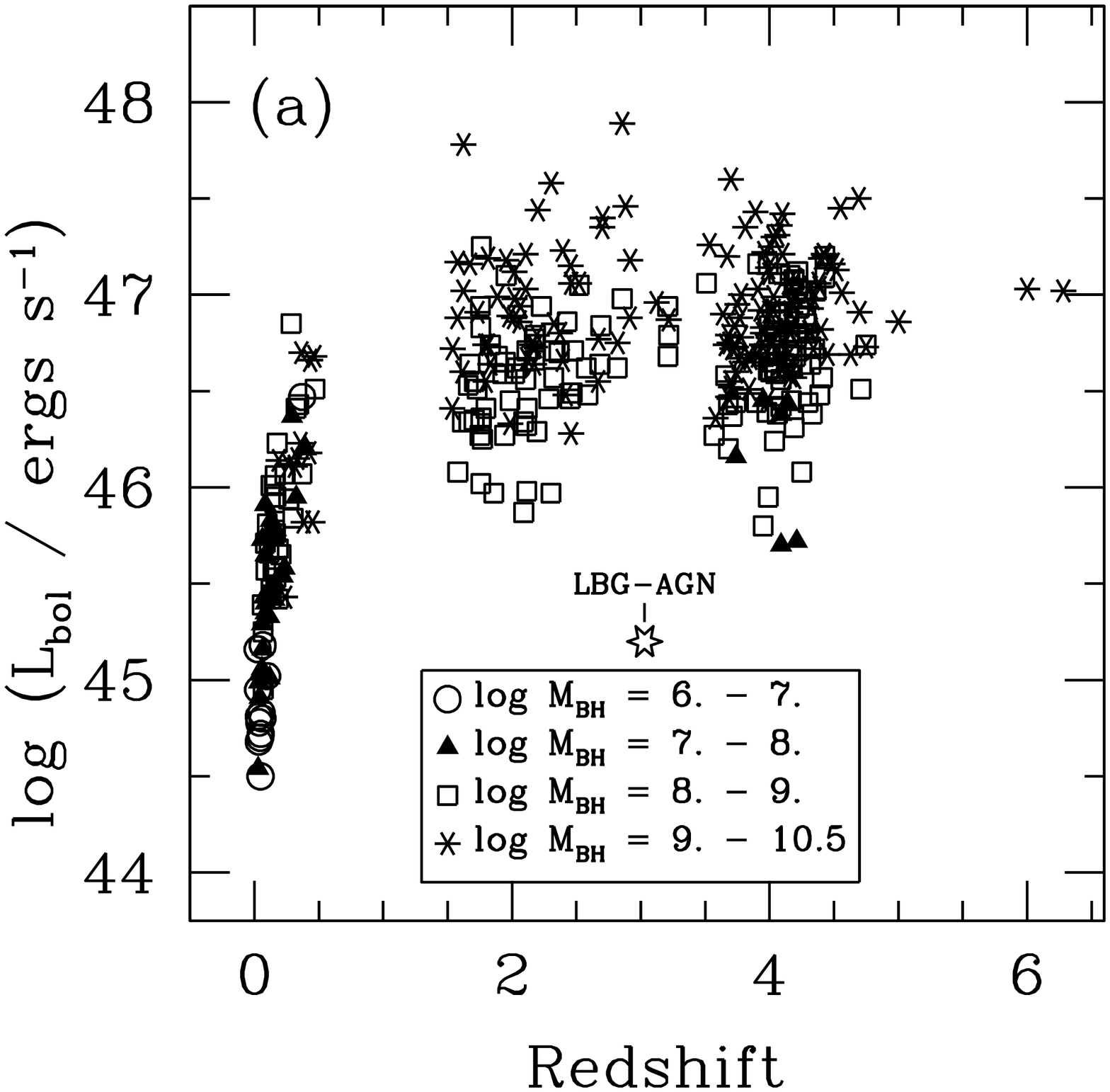}{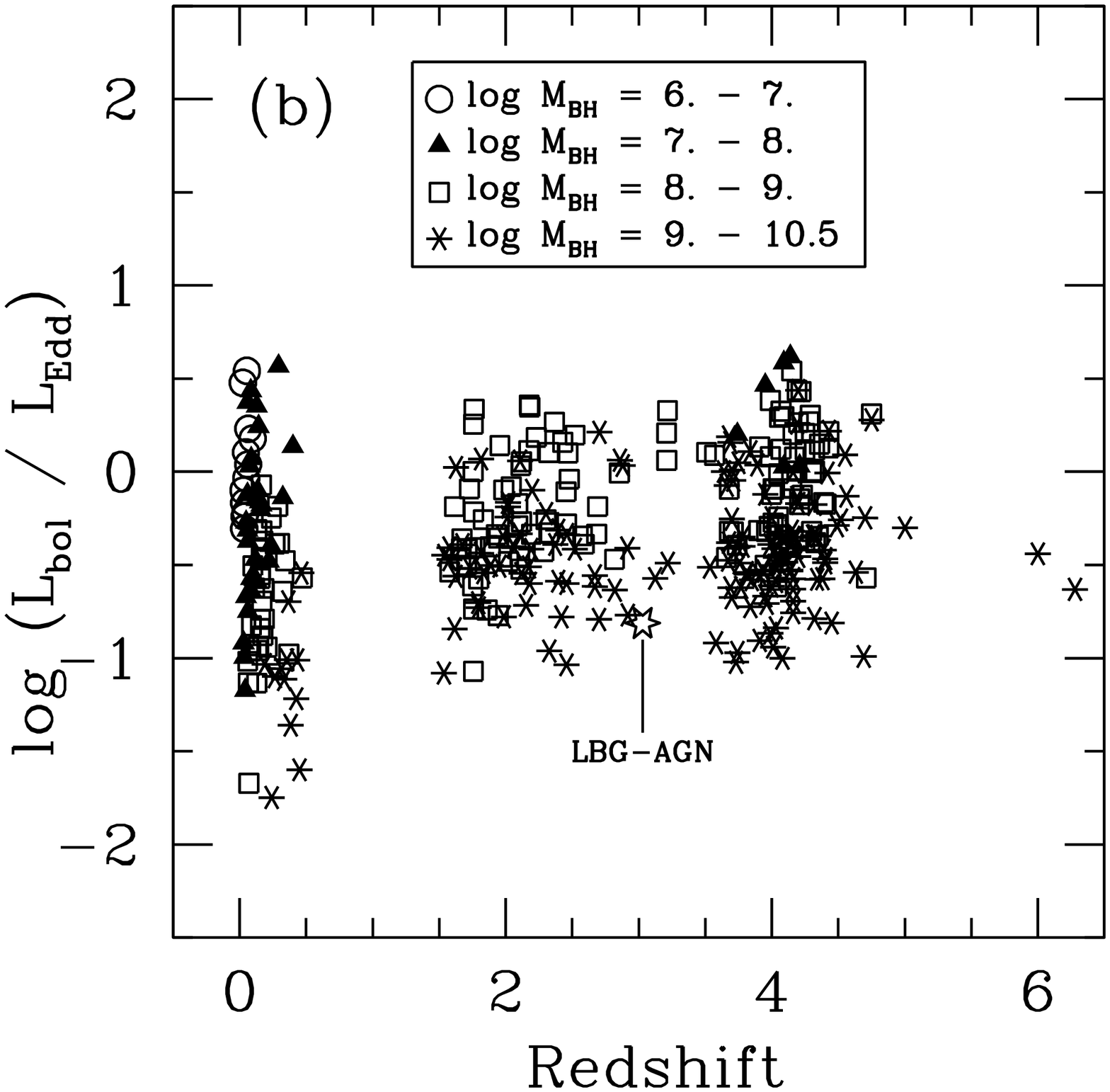}
\caption{Distributions of ($a$) \lbol{} and ($b$) \lol{} with redshift, shown 
binned in central mass, \mbh. Errorbars are omitted for clarity.}
\end{figure}

\section{Black-Hole Masses of Distant, Luminous Quasars} 

Figure~1 shows the distribution of estimated \mbh{} values for
several samples of quasars: 87 objects from the Bright Quasar Survey at 
$z \leq 0.5$, a sample of $1.5 < z \lsim 3.5$ quasars, 
and recently published samples of $z \approx 4$ quasars, including SDSS
EDR data (\eg Anderson \et 2001; Fan \et 2001; Constantin \et 2002).
A cosmology with $H_0$ = 75 ${\rm km~ s^{-1} Mpc^{-1}}$, q$_0$ = 0.5, and
$\Lambda$ = 0 is used. 
The reader is referred to Paper~I for further details. 
For these samples Figure~2a shows the distributions of the bolometric 
luminosities, \lbol, determined by applying bolometric correction factors
to measured monochromatic continuum luminosities.
The lower envelopes in both the \lbol{} and \mbh{} values are due to the
flux limits in the sample selection or in the original surveys. 
Figure~2b displays the redshift distribution of the Eddington luminosity
ratio of the samples. However, the uncertainty in the \lol{} values 
is significant ($\gsim$0.7\,dex) due to the need to apply an average bolometric
correction to obtain \lbol{} and due to the accuracy of the \mbh{}
estimates (1$\sigma$\,error is\,$\sim$0.5\,dex relative to reverberation masses). 
The observed range in \lol{} values is the result of selection flux limits
(lower envelope) and the combination of a minimum line-width adopted ($\sim$2000\,km/s)
to identify quasars and the observed upper \lbol{} envelope. 
 Nonetheless, these crude \lol{} estimates show that
massive, high-$z$ quasars have Eddington ratios of order 0.1\,$-$\,1.0, as
commonly expected, as opposed to super-Eddington or \mbox{highly 
sub-Eddington ratios.  The main conclusions are:}
\begin{itemize}
\item
Distant, luminous quasars have very massive black holes, even at the highest
redshifts. In other words, black holes with very large masses, 
\mbh{} $\approx 10^9$\Msol{}, exist at large redshifts --- beyond the space density
drop at $z \approx 3$ and even as early as $z \approx$ 6.

\item
There appears to be a real physical ceiling of \mbh{} $\approx 10^{10}$\Msol{}
and \lbol{} $\approx 10^{48}$\ergs. This is partly owing to the steep cutoff
in the upper end of the luminosity (and mass) function. In addition, this
may signify a maximum sustainable mass and luminosity.

\item
The early appearance of supermassive black holes suggests that they form
early or fast. In fact, they may even form faster than the stars in the host 
galaxy as suggested by studies of high-$z$ host galaxies (see below). 
\end{itemize}

The notion that active massive black holes at high-$z$ may reside in
relatively young galaxies is based on largely circumstantial evidence that 
host galaxies of AGNs at $z \gsim$3 do not typically appear fully formed or 
have old stellar populations (see Paper~I for a more detailed discussion): 
(a) the presence of large masses
of cold molecular gas in high-$z$ hosts (\ie significant fuel exists for
future star formation) which for some high-$z$ sources are clearly separated
from the AGN nucleus or is distinctly extended (\eg Ohta \et 1996; 
Papadopoulos \et 2000; Carilli \et 2002, 2003), 
(b) inferred high star-formation rates from the far-IR luminosities in \mbox{high-$z$ AGNs 
(\eg Carilli \et 2001, 2003; Omont \et} 2001), 
(c) the presence of large-scale young stellar populations confirmed in at least
one well-studied $z \approx 4$ radio galaxy with high far-IR luminosity,
4C\,41.17 (\eg Dey \et 1997), 
(d) the strikingly different morphology of host galaxies of high-$z$ radio-quiet
quasars and radio galaxies compared to their lower-$z$ cousins: radio galaxies
exhibit multiple components embedded in large-scale diffuse emission indicative
of non-relaxed systems in the process of forming 
(\eg van Breugel \et 1998) 
while radio-quiet quasars appear to have smaller scale-lengths and a fraction of 
their final stellar mass (\eg Ridgway \et 2001; Papovich \et 2001), and
(e) the presence of massive, active black holes ($\sim10^8$\Msol; marked `LBG-AGN'
in Figures~1 and 2) in Lyman-break galaxies at $z \geq 3$; these are 
typically small, young starforming galaxies (Steidel \et 2002).
Rix \et (2001) and Omont \et (2001) provide independent arguments in
favor of a faster build-up of black-hole mass \mbox{compared to the stellar
mass in high-$z$ host galaxies.}

\end{document}